\documentclass[twocolumn,showpacs,preprintnumbers,amsmath,amssymb]{revtex4}
\usepackage{graphicx}
\usepackage{dcolumn}
\usepackage{bm}

\begin{document}

\title{Longitudinal and Transverse Form Factors from $^{12}$C\,\,\footnote{\,\,Support by Conselho Nacional de Desenvolvimento Cient\'{\i}fico e
Tecnol\'{o}gico (CNPq) (Brazil), and the Third World Academy of
Science (TWAS) (Italy) for under grant of the scheme (TWAS-CNPq
exchange programs for postdoctoral researchers).}}

\author{F.~A.~Majeed$^{1,\footnote{\,\,Permenant address: Department of Physics, College of Science, Al-Nahrain University, Baghdad, IRAQ. \,\,\,Email:fouad@if.ufrj.br}}$}

\address{$^1$Instituto de F\'{\i}sica, Universidade Federal do
Rio de Janeiro, C.P. 68528, 21941-972 Rio de Janeiro, RJ, Brazil}

\date{\today}

\begin{abstract}
Electron scattering form factors from $^{12}$C have been studied in
the framework of the particle-hole shell model. Higher
configurations are taken into account by allowing particle-hole
excitations from the 1$s$ and 1$p$ shells core orbits up to the
1$f$-2$p$ shell. The inclusion of the higher configurations modifies
the form factors markedly and describes the experimental data very
well in all momentum transfer regions.
\end{abstract}

\maketitle

\section{Introduction}
Shell model calculations, carried out within a model space in which
the nucleons are restricted to occupy a few orbits are unable to
reproduce the measured static moments or transition strengths
without scaling factors. Inadequacies in the shell model
wavefunctions are revealed by the need to scale the matrix elements
of the one-body operators by effective charges to match the
experimental data. However, the introduction of effective charges
may bring the calculated transition strengths which are defined at
the photon point, as well as, the form factors at the first maximum,
closer to the measured values, but the non-zero momentum transfer
($q$) values might deviate appreciably from the measured values
\cite{RA03}.

Electron scattering at 200 MeV on $^{12}$C and $^{13}$C, have been
studied by Sato {\em et al.} \cite{TS85}. The effect of higher
configurations wavefunctions are included in the work of Bennhold
{\em et al.} \cite{CB85}. Booten {\em et al.} \cite{JB94}
investigated the higher configurations contributions on some
$p$-shell nuclei. Coulomb form factors of C2 transitions in several
selected $p$-shell nuclei are discussed by Radhi {\em et al.}
\cite{RA01} taking into account core-polarization effects.
Configuration mixing shell model has been recently used \cite{FA06}
to study the isovector states of $^{12}$C in the framework of
particle-hole theory. The calculations are quite successful and
describe the experimental form factors very well for all momentum
transfer regions.

The purpose of the present work is to include higher-energy
configurations by allowing excitation from 1$s$ and 1$p$ shells core
orbits up to the 1$f$-2$p$ shell. The configurations which include
the higher configurations is called the extended space
configurations. The ground state of $^{12}$C is taken to have closed
1$s$$_{1/2}$ and 1$p$$_{1/2}$ shells. The states expected to be most
strongly excited from closed-shell nuclei are linearly combination
of a configurations in which one nucleon has been raised to a higher
shell, forming pure single-particle-hole state \cite{TW84}. This
approximation is called Tamm-Dancoff approximation (TDA)\cite{TJ66}.
The dominant dipole, quadrupole and multipole $T$=1  single
particle-hole states of $^{12}$C are considered with the framework
of the harmonic oscillator (HO) shell model. The Hamiltonian is
diagnoalized in the space of the single-particle hole states, in the
presence of the modified surface delta interaction (MSDI)
\cite{PM77}. The space of the single-particle-hole states include
all shells up to 2$p$$_{1/2}$ shell. Admixture of higher
configurations is also considered. A comparison of the calculated
form factors using this model with the available experimental data
for the dominantly $T$=1 states are discussed.

\section{Theory}
The ground state of $^{12}$C is taken to have closed 1$s$$_{1/2}$
and 1$p$$_{3/2}$ shells, and is represented by $\Psi$$_{0}$. The
particle-hole state formed by promoting one particle from the
shell-model ground state. The particle-hole state of the total
Hamiltonian is represented by $\Phi$$_{JM}$($ab$$^{-1}$) with labels
(a) for particles with quantum numbers
($n$$_{a}$$\ell$$_{a}$$j$$_{a}$) and (b) for holes with quantum
numbers ($n$$_{b}$$\ell$$_{b}$$j$$_{b}$). The state
$\Phi$$_{JM}$($ab$$^{-1}$) indicating that a particle was vacated
from $j$$_{b}$ and promoted to $j$$_{a}$.

The excited state wavefunction can be constructed as a linear
combinations of pure basis  $\Phi$$^{,s}$  as \cite{TW84}
\begin{equation}
\Psi^{n}_{JM}=\sum_{ab}\chi^{J}_{ab^{-1}}\Phi_{JM}(ab^{-1}),
\end{equation}
where the amplitude $\chi$$^{J}_{ab^{-1}}$ can be determined from a
diagonalization of the residual interaction. By including the
isospin $T$ \cite{TJ66}, one now has to solve the secular equation
\begin{equation}
\sum_{ab}[\langle{\acute{a}\acute{b}^{-1}}|H|ab^{-1}\rangle_{JMTT_z}-E_n\delta_{\acute{a}\acute{b}^{-1},
ab^{-1}}]\,\chi^{JT}_{ab^{-1}}=0.
\end{equation}
The matrix element of the Hamiltonian is given by \cite{PM77}
\begin{eqnarray}
\langle{\acute{a}\acute{b}^{-1}}|H|ab^{-1}\rangle_{JMTT_z}
=(e_{\acute{a}}-e_{\acute{b}})\,\delta_{{a\acute{a}},{b\acute{b}}}\nonumber\\
+\langle{\acute{a}\acute{b}^{-1}}|V|ab^{-1}\rangle_{JMTT_z},
\end{eqnarray}
where $e$$_{\acute{a}}$-$e$$_{\acute{b}}$ is the unperturbed energy
of the particle-hole pair obtained from energies in nuclei with
A$\pm$1 particles.

The matrix element of the residual interaction $V$ is given by the
MSDI with the strength parameters $A$$_{0}$=0.8 MeV, $A$$_{1}$=1.0
MeV, $B$=0.7 MeV and $C$=$-$0.3 MeV \cite{PM77}.
\begin{eqnarray}
\langle{\acute{a}\acute{b}^{-1}}|V|ab^{-1}\rangle_{JMTT_z}
=-\sum_{\acute{J}\acute{T}}(2\acute{J}+1)(2\acute{T}+1)\nonumber\\\times\left\{
\begin{array}{ccc}
j_{\acute{a}} &  j_{b} & \acute{J}\\
j_{a} & j_{\acute{b}} & J
\end{array}\right\}\left\{
\begin{array}{ccc}
\frac{1}{2} &  \frac{1}{2} & T\\
\frac{1}{2} & \frac{1}{2} & \acute{T}
\end{array}\right\}\langle{\acute{a}\,b}|V|a\acute{b}\rangle_{\acute{J}\,\acute{T}}.
\end{eqnarray}

The matrix elements of the multipole operators $T$$_{J}$ are given
in terms of the single particle matrix elements by \cite{TW84}
\begin{equation}
     \left\langle\Psi_{J}\|T_{Jt_{z}}\|\Psi_{0}\right\rangle
     =\sum_{ab}\chi^{Jt_{z}}_{ab^{-1}}\left\langle
     a\|T_{Jt_{z}}\|b\right\rangle,
\end{equation}
where $t$$_{z}$=1/2 for protons and -1/2  for neutrons. The
amplitudes $\chi$$^{Jt_{z}}_{ab^{-1}}$ can be written in terms of
the amplitudes $\chi$$^{JT}_{ab^{-1}}$ in isospin space as
\cite{PM77}
\begin{eqnarray}\
      \chi^{Jt_{z}}_{ab^{-1}}=(-1)^{T_{f}-T_{i}}\left[\left(
                                                        \begin{array}{ccc}
                                                          T_{f} & 0 & T_{i} \\
                                                          -T_{z} & 0 & T_{z} \\
                                                        \end{array}
                                                      \right)\sqrt{2}\ \frac{\chi^{JT=0}_{ab^{-1}}}{2}\right.\nonumber\\
      \left.+2t_{z}\left(\begin{array}{ccc}
                  T_{f} & 0 & T_{i} \\
                  -T_{z} & 0 & T_{z} \\
                  \end{array}
                  \right)\sqrt{6}\
                  \frac{\chi^{JT=1}_{ab^{-1}}}{2}\right],
\end{eqnarray}
where
\begin{equation}\
    T_{z}=\frac{Z-N}{2}
\end{equation}

The single particle matrix elements of the electron scattering
operator $T$$^{\eta}_{J}$ are those of Ref.\cite{BA85} with $\eta$
selects the longitudinal ($L$), transverse electric ($E\ell$) and
transverse magnetic ($M$) operators, respectively. Electron
scattering form factors involving angular momentum transfer $J$ is
given by \cite{BA85}
\begin{eqnarray}\
    |F^{\eta}_{J}(q)|^{2}=\frac{4\pi}{Z^{2}(2J_{i}+1)}\ |\langle\Psi_{J_{f}}\|T^{\eta}_{Jt_{z}}\|\Psi_{J_{i}}\rangle\nonumber\\
    \times|F_{c.m}(q)|^{2} \ |F_{f.s}(q)|^{2}
\end{eqnarray}
where $J$$_{i}$= 0 and $J$$_{f}$=$J$ for closed shell nuclei and $q$
is the momentum transfer. The last two terms in Eq.\,(8) are the
correction factors for the ($c.m.$) and the finite nucleon size
($f.s.$)\cite{BA85}. The total inelastic electron scattering form
factor is defined as \cite{TJ66}
\begin{equation}
     |F_{J}(q,\theta)|^{2}=|F^{L}_{J}(q)|^{2}+\left[\frac{1}{2}+\tan^{2}(\theta/{2})\right]
     |F^{Tr}_{J}(q)|^{2},
\end{equation}
where $|F^{Tr}_{J}(q)|^{2}$ is the transverse electric or transverse
magnetic form factors.

\section{Results and Discussion}
The unperturbed energies for the single particle-hole states for
both positive and negative parity states used in this work are
adopted from our previous theoretical work (see Table 1 and 2 from
Ref.\cite{FA06}). Higher configurations are included in the
calculations when the ground state is considered as a mixture of the
$|(1$$s$$_{1/2}$)$^{4}$$\,(1$$p$$_{3/2}$)$^{8}$$\rangle$ and
$|(2$$s$$_{1/2}$)$^{4}$$\,(2$$p$$_{3/2}$)$^{8}$$\rangle$
configurations, such that the ground state wavefunction becomes
\begin{eqnarray}
|\Psi_{00}\rangle=\gamma|\Psi_{00}(1s_{1/2})^4(1p_{3/2})^8\rangle\nonumber\\
                                        +\delta|\Psi_{00}(2s_{1/2})^4(2p_{3/2})^8\rangle
\end{eqnarray}

with $\gamma^{2}$+$\delta^{2}$=1,
$\chi^{JT}_{ab_{1}^{-1}}$=$\gamma\chi^{JT}_{ab^{-1}}$ and
$\chi^{JT}_{ab_{2}^{-1}}$=$\delta\chi^{JT}_{ab^{-1}}$

The excited states is also assumed as a mixture of the particle-hole
configurations, $|$a$_{1}$\,$b^{-1}_{1}$$\rangle$,
$|$a$_{2}$\,$b^{-1}_{2}$$\rangle$, $|$a$_{2}$\,$b^{-1}_{1}$$\rangle$
and $|$a$_{1}$\,$b^{-1}_{2}$$\rangle$, where
$|$a$_{1}$$\rangle$=$|$a$\rangle$=$|$$n$$_{a}$$\,$$\ell$$_{a}$$\,$$j$$_{a}$$\rangle$,
$|$a$_{2}$$\rangle$=$|$a$\rangle$=$|$$n$$_{a}$$+1\,$$\ell$$_{a}$$\,$$j$$_{a}$$\rangle$,
$|$b$_{1}$$\rangle$=$|$b$\rangle$=$|$$n$$_{b}$$\,$$\ell$$_{b}$$\,$$j$$_{b}$$\rangle$
and
$|$b$_{2}$$\rangle$=$|$b$\rangle$=$|$$n$$_{b}$$+1\,$$\ell$$_{b}$$\,$$j$$_{b}$$\rangle$.

The matrix element given in Eq.\,(5) is called the model space
matrix element, where $a$ and $b$ are defined by the amplitudes
given in Tables~\ref{tab1} and \ref{tab2} for the negative and
positive parity states, respectively.

The extended space matrix element becomes

\begin{eqnarray}
     \left\langle\Psi_{J}\|T_{Jt_{z}}\|\Psi_{0}\right\rangle
     =\sum_{a_{1}b_{1}}\chi^{Jt_{z}}_{a_{1}b_{1}^{-1}}\left\langle
     a_{1}\|T_{Jt_{z}}\|b_{1}\right\rangle\nonumber\\
     +\sum_{a_{1}b_{2}}\chi^{Jt_{z}}_{a_{1}b_{2}^{-1}}\left\langle
     a_{1}\|T_{Jt_{z}}\|b_{2}\right\rangle\nonumber\\
     +\sum_{a_{2}b_{1}}\chi^{Jt_{z}}_{a_{2}b_{1}^{-1}}\left\langle
     a_{2}\|T_{Jt_{z}}\|b_{1}\right\rangle\nonumber\\
     +\sum_{a_{2}b_{2}}\chi^{Jt_{z}}_{a_{2}b_{2}^{-1}}\left\langle
     a_{2}\|T_{Jt_{z}}\|b_{2}\right\rangle,
\end{eqnarray}
where
\begin{eqnarray}
    \chi^{Jt_{z}}_{a_{1}b_{1}^{-1}}=C_{1}\,\chi^{Jt_{z}}_{ab^{-1}},\nonumber\\
    \chi^{Jt_{z}}_{a_{1}b_{2}^{-1}}=C_{2}\,\chi^{Jt_{z}}_{ab^{-1}},\nonumber\\
    \chi^{Jt_{z}}_{a_{2}b_{1}^{-1}}=C_{3}\,\chi^{Jt_{z}}_{ab^{-1}},\nonumber\\
    \chi^{Jt_{z}}_{a_{2}b_{2}^{-1}}=C_{4}\,\chi^{Jt_{z}}_{ab^{-1}},
\end{eqnarray}

\begingroup
\begin{table}
 \caption{\label{tab1} Energies and amplitudes $\chi$$^{JT}$ for  $J$$^{-}$ $T=1$ state.}
\begin{ruledtabular}
\begin{center}
\begin{tabular}{lcccc}

Particle-hole & E($1$$^{-}$) &  E($2_{1}$$^{-}$)& E($2_{2}$$^{-}$)&
E($3$$^{-}$)\\
configuration & 18.44 MeV & 19.88 MeV & 23.50 MeV & 18.87 MeV \\
$|$a\,$b^{-1}$$\rangle$ & $\chi$$^{11}$ & $\chi$$^{21}$ & $\chi$$^{31}$ & $\chi$$^{31}$\\
\hline \\
 (1$p$$_{1/2}$)\,(1$p$$_{3/2}$)$^{-1}$ & 0.0473  & 0.0000 & 0.0000 & 0.0000\\
 (1$d$$_{5/2}$)\,(1$s$$_{1/2}$)$^{-1}$ & -0.1810 & 0.8314 & 0.0703 & 0.9993\\
 (2$s$$_{1/2}$)\,(1$s$$_{1/2}$)$^{-1}$ & 0.9739  & 0.5430 & 0.0834 & 0.0000\\
 (1$d$$_{1/2}$)\,(1$s$$_{1/2}$)$^{-1}$ & 0.1333  & -0.1054& 0.9936 & 0.0318\\
 (1$f$$_{7/2}$)\,(1$p$$_{3/2}$)$^{-1}$ & 0.0000  & 0.0442 & 0.0000 & 0.0165\\
 (2$p$$_{3/2}$)\,(1$p$$_{3/2}$)$^{-1}$ & 0.0008  & 0.0000 & 0.0222 & 0.0000\\
 (1$f$$_{5/2}$)\,(1$p$$_{3/2}$)$^{-1}$ & 0.0000  & -0.0636& 0.0147 &-0.0030\\
 (2$p$$_{1/2}$)\,(1$p$$_{3/2}$)$^{-1}$ & 0.0000  & 0.0000 & 0.0000 & 0.0000\\
\end{tabular}
\end{center}
\end{ruledtabular}
\end{table}
\endgroup

\begingroup
\begin{table}
 \caption{\label{tab2} Energies and amplitudes $\chi$$^{JT}$ for $J$$^{+}$ $T=1$ states.}
\begin{ruledtabular}
\begin{center}
\begin{tabular}{lc}

  Particle-hole &  E($3$$^{+}$)=27.10 MeV \\
  configuration &  $\chi$$^{31}$\\
  $|$a\,$b^{-1}$$\rangle$ & \\
\hline \\
 (1$p$$_{1/2}$)\,(1$p$$_{3/2}$)$^{-1}$ & 0.0000\\
 (1$d$$_{5/2}$)\,(1$s$$_{1/2}$)$^{-1}$ & -0.0475\\
 (2$s$$_{1/2}$)\,(1$s$$_{1/2}$)$^{-1}$ & 0.0000\\
 (1$d$$_{1/2}$)\,(1$s$$_{1/2}$)$^{-1}$ & 0.0000\\
 (1$f$$_{7/2}$)\,(1$p$$_{3/2}$)$^{-1}$ & 0.9461\\
 (2$p$$_{3/2}$)\,(1$p$$_{3/2}$)$^{-1}$ & -0.3201\\
 (1$f$$_{5/2}$)\,(1$p$$_{3/2}$)$^{-1}$ & -0.0020\\
 (2$p$$_{1/2}$)\,(1$p$$_{3/2}$)$^{-1}$ & 0.0000 \\
\end{tabular}
\end{center}
\end{ruledtabular}
\end{table}
\endgroup

The values of the parameters C$^{,s}$ are given in Table~\ref{tab3}.
The states $1$$^{-}$, $2_{1}$$^{-}$, $2_{2}$$^{-}$, $3$$^{-}$ and
$3$$^{+}$ are found experimentally at 18.12 MeV, 19.50 MeV, 22.70
MeV, 18.60 MeV and 20.60 MeV respectively \cite{RR87}. We obtain the
values 18.44 MeV, 19.88 MeV, 23.50 MeV, 18.87 MeV and 27.10 MeV for
the states $1$$^{-}$, $2_{1}$$^{-}$, $2_{2}$$^{-}$, $3$$^{-}$ and
$3$$^{+}$, respectively.

\begingroup
\begin{table}
 \caption{\label{tab3} Values of the parameters $C$$^{,s}$  used in the extended space calculations.}
\begin{ruledtabular}
\begin{center}
\begin{tabular}{ccccc}

  $J$$^{\pi}$ &  $C$$_{1}$ & $C$$_{2}$ & $C$$_{3}$ & $C$$_{4}$\\
 \hline \\
 3$^{+}$ & 0.92 & -0.27 & -0.27 & 0.078 \\
 2$_{1}^{-}$ &  -0.92 & 0.27 & -0.27 & 0.078 \\
 2$_{2}^{-}$ &  -0.92 & 0.27 & -0.27 & 0.078 \\

\end{tabular}
\end{center}
\end{ruledtabular}
\end{table}
\endgroup
The $1$$^{-}$ (18.12 MeV), C1+E1 form factor is shown in Fig.\,1.
The amplitudes $\chi$$^{,s}$ reduced by a factor 1.3, to agree with
the low $q$ data \cite{TW84}. This state is dominated by
(2$s$$_{1/2}$)\,(1$s$$_{1/2}$)$^{-1}$ particle-hole configuration,
as given in Table~\ref{tab1}. The single-particle matrix elements
are calculated with the harmonic oscillator wavefunctions (HO) with
oscillator parameter $b=1.64$\, fm  to agree with the elastic form
factor determination \cite{TK85}. Our results are consistent with
the previous calculation of Donnelly \cite{TW70} and slightly in
better agreement with the experimental data for the momentum
transfer region $q \leq$ 1.0 fm$^{-1}$.

The transverse magnetic form factor M2 for the excitation to the
$2_{1}$$^{-}$, 19.50 MeV state is shown in Fig.\,2. The amplitudes
have to be enhanced by a factor 1.2 to account for the experimental
data. The calculations incorporate the single-particle wavefunctions
of the (HO) potential with $b=1.64$\, fm and a value of
$\gamma$=0.95, to account for the ground state correlation. The data
are very well explained for the momentum-transfer $q \leq$ 3.0
fm$^{-1}$.

Figure\,3, shows the transverse magnetic form factor M2 for the
excitation to the $2_{1}$$^{-}$, 22.70 MeV state. The amplitudes
have to be reduced by a factor 1.82 to fit the low-$q$ data. The
single-particle wavefunctions are those of the (HO) potential with
size parameter $b=1.50$\, fm and a value of $\gamma$=0.97, to
account for the ground state correlation. The experimental data are
very well described throughout the momentum-transfer regions and the
results are consistent with that of Hicks {\em et al.}, \cite{RS84}.

The $3$$^{-}$ (18.60 MeV), is dominated by
(1$d$$_{5/2}$)\,(1$s$$_{1/2}$)$^{-1}$ particle-hole configuration,
as given in Table~\ref{tab1}. The only multipole that contributes to
the scattering is the  longitudinal C3 multipole as shown in
Fig.\,4. The calculations incorporate the single-particle
wavefunctions of the (HO) potential with $b=1.64$\, fm and $\gamma$
takes the value 1.0 . The experimental data are very well explained
for the momentum-transfer values $q \leq$ 3.0 fm$^{-1}$ and the
results are consistent with that of Hicks {\em et al.}, \cite{RS84}
and Yamaguchi {\em et al.}, \cite{YM71}, where the form factor seems
to be a pure longitudinal form factor.

Figure\,5, shows the transverse magnetic form factor for the
excitation to the $3$$^{+}$, 20.60 MeV state. The dominated
configuration is the (1$f$$_{7/2}$)\,(1$p$$_{3/2}$)$^{-1}$
particle-hole configuration, as given in Table~\ref{tab2}. The only
multipole that contributes to the scattering is the magnetic M3
multipole. The amplitudes have to be reduced by factor of 5 to
account for the experimental data. The calculations incorporate the
single-particle wavefunctions of the (HO) potential with $b=1.64$\,
fm, and a value $\gamma$=0.7, to account for the ground state
correlation. The data are very well explained throughout the
momentum-transfer values $q \leq$ 3.0 fm$^{-1}$.

\begin{figure}
\centering
\includegraphics[width=0.4\textwidth]{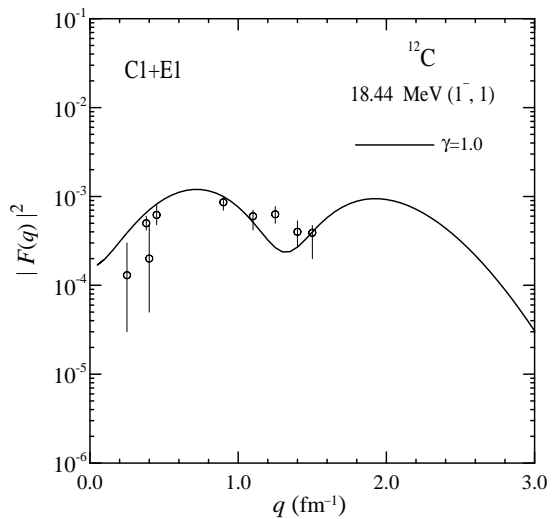}
\caption{Form factor for the C1+E1 transition to the ($1$$^{-}$, 1)
18.44 MeV state compared with the experimental data taken from Ref.
\cite{TW70}.}
\end{figure}

\begin{figure}
\centering
\includegraphics[width=0.4\textwidth]{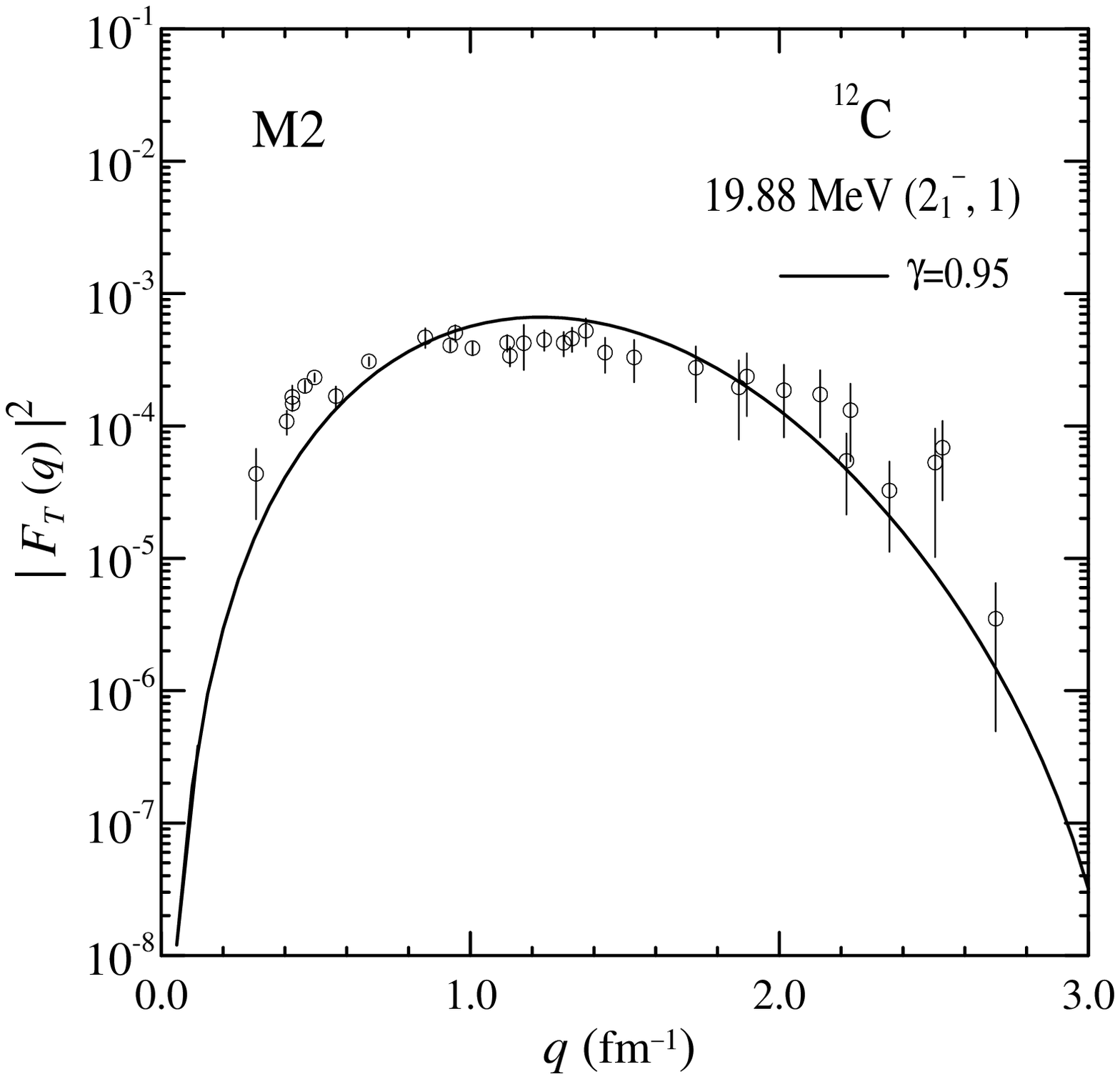}
\caption{Transverse magnetic form factor for the M2 transition to
the ($2_{1}$$^{-}$, 1) 19.88 MeV state compared with the
experimental data taken from Ref. \cite{RS84}.}
\end{figure}

\begin{figure}
\centering
\includegraphics[width=0.4\textwidth]{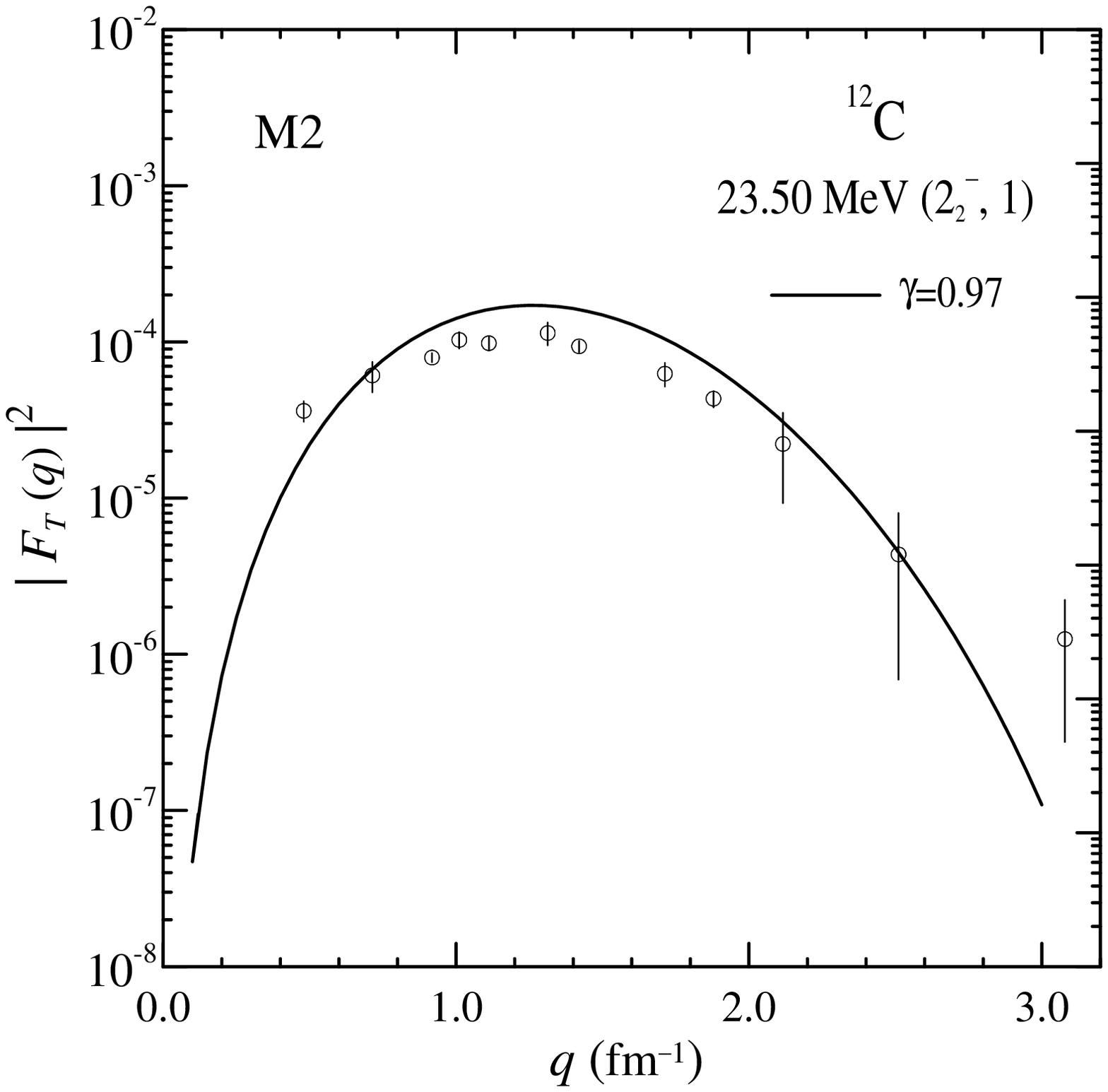}
\caption{Transverse magnetic form factor for the M2 transition to
the ($2_{2}$$^{-}$, 1) 23.50 MeV state compared with the
experimental data taken from Ref. \cite{RR87}.}
\end{figure}

\begin{figure}
\centering
\includegraphics[width=0.4\textwidth]{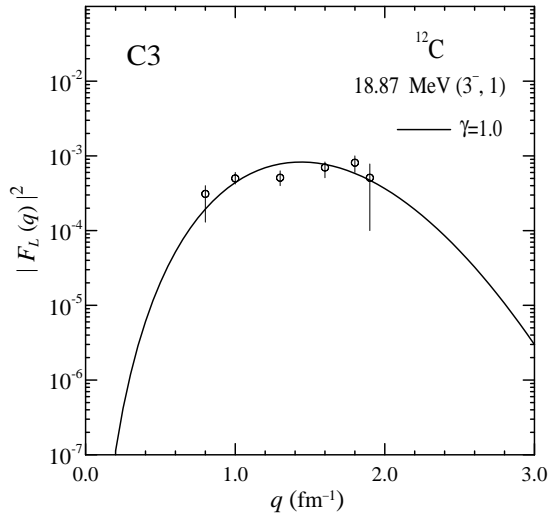}
\caption{Longitudinal form factor for the C3 transition to the
(3$^{-}$, 1) 18.87 MeV state compared with the experimental data
taken from Ref. \cite{YM71}.}
\end{figure}

\begin{figure}
\centering
\includegraphics[width=0.4\textwidth]{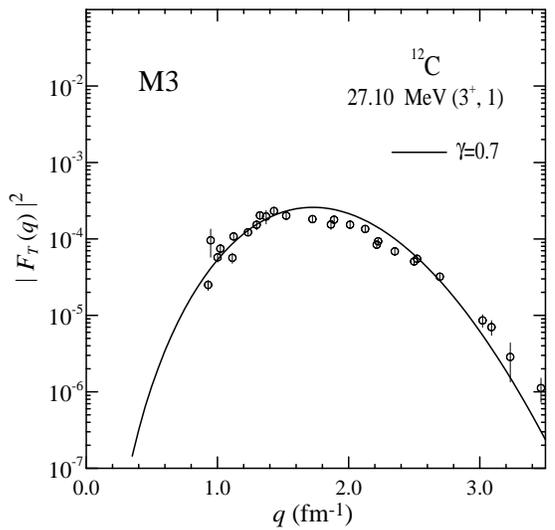}
\caption{Transverse magnetic form factor for the M3 transition to
the (3$^{+}$, 1) 27.10 MeV state compared with the experimental data
taken from Ref. \cite{RS84}.}
\end{figure}

\newpage
\section{Conclusions}
The inclusion of higher energy configurations in the particle-hole
shell model calculation succeeded in describing the form factors for
the negative and positive parity states. The  amplitudes of the
transitions to the negative-parity states considered in this work
have to be reduced by a factor  1.3 and 1.82 for the states
$1$$^{-}$ and $2_{2}$$^{-}$ while the amplitudes for the
$2_{1}$$^{-}$ state need to be enhanced by factor of 1.2, to
describe the low-q data. The amplitudes for $3$$^{+}$ need to be
reduced by a factor of 5. This reduction may be attributed to higher
order effects, such as 2p-2h excitations, or even more. Correlation
in the ground state wavefunction by mixing more than one
configuration are necessary to describe the data. The
single-particle wavefunctions of the (HO) potential with size
parameter $b=1.64$\, fm chosen to reproduce the root mean square
charge radius are adequate to describe the data, except for M2
(23.50 MeV) transition where the $b$ value has to be reduced by a
factor 14\%.

\newpage

\begin{thebibliography}{99}

\bibitem{RA03}R.~A.~Radhi,~A.~Bouchebak,~Nucl.~Phys.~A{\bf~716},~87~(2003).
\bibitem{TS85}T.~Sato,~{\em et al.},~Z.Phys.~A{\bf~320},~507~(1985).
\bibitem{CB85}C.~Bennhold,~{\em et al.},~Phys.~Rev.~C{\bf~46},~2456~(1992).
\bibitem{JB94}J.~G.~L.~Booten,~{\em et al.},Nucl.~Phys.~A{\bf~569},~510~(1994).
\bibitem{RA01}R.~A.~Radhi,~{\em et al.},Nucl.~Phys.~A{\bf~696},~442~(2001).
\bibitem{FA06}F.~A.~Majeed,~R.~A.~Radhi,~Chin.~Phys.~Lett.~Vol.~{\bf23},
No.10,~2699 (2006).
\bibitem{TW84}T.~W.~Donnelly,~I.~Sick,~Rev.~Mod.~Phys.~Vol.~{\bf 56},~(3),~461
(1984).
\bibitem{TJ66}T.~deForest,~Jr.,~J.~D.~Walecka,~ Adv.~Phys.~{\bf 15},~1
(1966).
\bibitem{PM77}P.~J.~Brussaard,~P.~W.~M.~Glaudemans,~{\em Shell-Model
              Applications  in  Nuclear  Spectrscopy} (Amsterdam: North
              Holland),(1977).
\bibitem{BA85}B.~A.~Brown,~{\em et al.},~Phys.~Rev.~C{\bf
32},~1127~(1985).
\bibitem{RR87}R.~S.~Hicks,~{\em etal.},~Phys.~Rev.~C{\bf~36},~485~(1987).
\bibitem{TW70}T.~W.~Donnelly,~Phys.~Rev.~C{\bf~1},~833~(1970).
\bibitem{TK85}T.~Sato,~{\em et al.},~Z.~Phys.~A{\bf 320},~507
(1985).
\bibitem{RS84}R.~S.~Hicks,~{\em et al.},~Phys.~Rev.~C{\bf~30},~1~(1984).
\bibitem{YM71}A.~Yamaguchi,~{\em
et al.},~Phys.~Rev.~C{\bf~3},~1750~(1971).

\end{thebibliography}

\end{document}